\documentclass[11pt,a4paper]{article}
\pdfoutput=1
\usepackage{jcappub}

\usepackage{mathrsfs}
\usepackage{bm}
\usepackage{subfigure}
\usepackage{feynmp}
\usepackage{extarrows}
\usepackage{slashed}
\usepackage{dcolumn}% Align table columns on decimal point
\usepackage{verbatim}
\usepackage{here}
\usepackage{multirow} %For Tabular.
\usepackage{xcolor}
\allowdisplaybreaks[4]
\numberwithin{equation}{section}

\newcommand{\FR}[2]{\displaystyle\frac{\,{#1}\,}{#2}}
\newcommand{\fr}[2]{\mbox{$\frac{\,{#1}\,}{#2}$}}

\def\bge{\begin{equation}}
\def\ede{\end{equation}}
\def\bga{\begin{aligned}}
\def\eda{\end{aligned}}
\newcommand{\beq}{\begin{equation}}
\newcommand{\eeq}{\end{equation}}
\newcommand{\bq}{\begin{equation}}
\newcommand{\eq}{\end{equation}}
\newcommand{\ba}{\begin{array}}
\newcommand{\ea}{\end{array}}
\newcommand{\beqa}{\begin{eqnarray}}
\newcommand{\eeqa}{\end{eqnarray}}
\newcommand{\beqs}{\begin{subequations}}
\newcommand{\eeqs}{\end{subequations}}

\def\({\left(}
\def\){\right)}

\def\End{\end{document}}
\newcommand{\order}[1]{\mathcal{O}({#1})}
\def\bge{\begin{equation}}
\def\ede{\end{equation}}
\def\bga{\begin{aligned}}
\def\eda{\end{aligned}}
\def\bgp{\begin{pmatrix}}
\def\edp{\end{pmatrix}}
\def\bgs{\begin{subequations}}
\def\eds{\end{subequations}}

\def\leqq{\leqslant}

\def\al{\alpha}
\def\be{\beta}

\def\la{\lambda}

\def\ga{\gamma}
\def\de{\delta}
\def\ep{\epsilon}
\def\lam{\lambda}
\def\rh{\rho}
\def\si{\sigma}

\def\bge{\begin{equation}}
\def\ede{\end{equation}}
\def\bga{\begin{aligned}}
\def\eda{\end{aligned}}
\def\bgp{\begin{pmatrix}}
\def\edp{\end{pmatrix}}
\def\bgs{\begin{subequations}}
\def\eds{\end{subequations}}
\def\di{{\mathrm{d}}}
\def\D{{\mathrm{D}}}

\def\TR{T_{\text{reh}}^{}}

\def\pd{\partial}
\def\ld{{\mathscr{L}}}

\def\la{\langle}\def\ra{\rangle}

\def\Tr{\mathrm{\,Tr\,}}

\def\to{\rightarrow}

\def\ii{\mathrm{i}}

\def\al{\alpha}
\def\be{\beta}
\def\ga{\gamma}
\def\de{\delta}
\def\ep{\epsilon}

\def\lam{\lambda}
\def\rh{\rho}
\def\si{\sigma}

\def\MP{M_{\text{P}}^{}}

\def\Mp{M_{\text{P}}}

\def\vG{v_G^{}}

\def\kp{\mathcal{K}}
\def\sp{\mathcal{W}}

\newcommand{\ob}[1]{\mkern 2mu \overline{\mkern -2mu #1 \mkern -2mu}\mkern 2mu}
\newcommand{\wt}[1]{\mkern 2mu \widetilde{\mkern -2mu #1 \mkern -2mu}\mkern 2mu}

\def\End{\end{document}}

\title{\huge Higgs\,Inflation,\,Reheating\,and\,Gravitino\,Production
in~No-Scale~Supersymmetric~GUTs}

\author[a]{\large John Ellis,}
\author[b]{\large~~Hong-Jian He,}
\author[c]{\large~~Zhong-Zhi Xianyu\,}

\affiliation[a\,]{Theoretical Particle Physics and Cosmology Group, Department of Physics,\\
                  \,King's College London, London WC2R 2LS, UK; \\
                  \,Theoretical Physics Department, CERN, CH-1211 Geneva 23, Switzerland.}
\affiliation[b\,]{Institute of Modern Physics and Center for High Energy Physics,\\
                  \,Tsinghua University, Beijing 100084, China;\\
                  \,Center for High Energy Physics, Peking University, Beijing 100871, China.}
\affiliation[c\,]{Center of Mathematical Sciences and Applications, and Department of Physics,\\
                  \,Harvard University, Massachusetts 02138, USA.}

\emailAdd{john.ellis@cern.ch, hjhe@tsinghua.edu.cn, xianyu@cmsa.fas.harvard.edu}

\abstract{
\\[1mm]
We extend our previous study of supersymmetric Higgs inflation in the context of
no-scale supergravity and grand unification, to include models based on the flipped SU(5)
and the Pati-Salam group. Like the previous SU(5) GUT model, these yield a class
of inflation models whose inflation predictions interpolate between those of the
quadratic chaotic inflation and Starobinsky-like inflation, while avoiding tension
with proton decay limits. We further analyse the reheating process in these models,
and derive the number of $e$-folds, which is independent of
the reheating temperature. We derive the corresponding predictions for the scalar tilt
and the tensor-to-scalar ratio in cosmic microwave background perturbations,
as well as discussing the gravitino production following inflation.
}

\keywords{\\[1mm]
Inflation, Supersymmetry and Cosmology, Particle Physics\,$-$\,Cosmology Connection
\\[4mm]
KCL-PH-TH-2016-33, CERN-TH-2016-134. \hfill
JCAP (2016), in Press [arXiv:1606.02202].
}

\begin{document}

\maketitle

\setlength{\baselineskip}{18pt}

\setcounter{page}{2}
\vspace*{10mm}
\section{Introduction}
\label{sec:1}
\vspace*{2mm}

The scalar Higgs boson $h$(125\,GeV) holds a unique position
in the Standard Model (SM) of particle physics. Via its interactions with
the spin-1 weak gauge bosons and the spin-$\frac{1}{2}$ fermions (quarks and leptons),
its vacuum expectation value generates the masses of SM particles.
On the other hand, cosmological inflation\,\cite{INF} postulates a scalar inflaton field
to drive near-exponential expansion of the very early Universe, whose quantum fluctuations
generate the observed perturbations in the cosmic microwave background (CMB),
and thereby large-scale structure. However, the identity of the inflaton is unknown so far.
It is natural to identify the inflaton with the SM Higgs boson as postulated in models
of Higgs inflation\,\cite{HiggsInf},
and seek possible tests from cosmological observations and collider measurements.
In this regard, Higgs inflation would be a truly economical and predictive mechanism for
the cosmological inflation, and provide a welcome direct link between the SM
and the early-Universe cosmology.

One problem of minimal Higgs inflation in the SM is the instability or
metastability of the Higgs potential at high scales\,\cite{INS}\footnote{%
In some non-Higgs inflation models, vacuum stability could be restored by adding a
non-minimal Higgs-gravity coupling (with radiative corrections) and assuming inflation
is driven by new physics not directly coupled to the SM \cite{xx}.}.
The resolution of this problem probably calls for
some new physics beyond the SM\,\cite{HX2014,EHX2014}, such as supersymmetry (SUSY),
which can help to stabilize the Higgs potential\,\cite{Ellis2001HiggsSUSY},
while also predicting a fairly light Higgs boson that can be identified as the 125\,GeV
scalar particle discovered at the LHC.
SUSY could also serve to control the magnitudes of quantum corrections
to the parameters of the inflationary potential\,\cite{Cries}.
Since the energy scale of inflation is around that of
SUSY gauge unification, we are tempted to embed Higgs inflation
into certain SUSY grand unified theory (GUT), within a supergravity framework.
It is particularly attractive to choose no-scale supergravity (SUGRA)\,\cite{no-scale},
since it emerges naturally from simple string compactifications\,\cite{witten},
and provides flat directions which are advantageous for
cosmological applications\,\cite{nos-app}.

We have given previously an explicit realization of a no-scale supersymmetric GUT scenario for inflation\,\cite{EHX2014},
in which Higgs inflation is realized in an SU(5) GUT embedded in
no-scale supergravity. In this model, the inflaton is identified as
the $D$-flat direction of the two Higgs doublets in the minimal supersymmetric extension of the SM (MSSM).
The predictions of this model for scalar and tensor perturbations in the
cosmic microwave background (CMB) interpolate between the predictions of quadratic chaotic inflation
and the Starobinsky model. An important feature of this model is that a fairly flat inflaton
potential can be achieved without introducing a non-minimal Higgs-gravity coupling
or imposing a shift symmetry on the K\"ahler potential,
since the no-scale supergravity structure of this model
provides the desired flat direction.
In this no-scale SUSY GUT approach to Higgs inflation,
all Higgs bosons couple minimally to gravity via the energy-momentum tensor,
without any non-minimal coupling between the Higgs fields and the Ricci scalar.
Hence our no-scale Higgs inflation approach differs in an essential way
from traditional SM Higgs inflation\,\cite{HiggsInf},
other SUSY and GUT extensions in the literature\,\cite{Xext}, and the Starobinsky-like
no-scale supergravity scenario proposed in\,\cite{ENO6}.

In this work, we first show how the structure of our previous no-scale SU(5) GUT inflationary model
can be generalized to other GUT models, specifically the flipped
SU(5)\,\cite{Antoniadis1987231,Ellis19881} and Pati-Salam\,\cite{PatiSalam} GUTs.
One motivation for this generalization is that the simplest SUSY SU(5) GUT is already tightly
constrained by experiments, particularly the limits on proton decay.
These cause tension with the construction of~\cite{EHX2014},
which requires the colored Higgs fields to have masses around
$10^{13}$\,GeV.\footnote{One way to avoid this problem
is to invoke non-minimal contributions to
the gauge kinetic function in supergravity, which could modify the
gauge unification condition and thus relax the proton decay bound~\cite{EHX2014}.}\,
In the current study, we show that the constructions with the flipped SU(5) model
or with the Pati-Salam model can disentangle the colored Higgs mass from the scale of inflation,
making model building much more flexible.
In addition, neither of the flipped SU(5) and Pati-Salam GUTs require any field
in an adjoint representation, so they can be embedded more naturally into string theory.
However, our constructions of the flipped SU(5) and Pati-Salam GUT Higgs inflation models make similar
predictions as the SU(5) no-scale model\,\cite{EHX2014}, also interpolating
between the quadratic and Starobinsky potentials.

After presenting the no-scale Higgs inflation models \`{a} la flipped SU(5) and Pati-Salam,
we investigate the reheating process in these GUT models.
Since reheating happens after inflation, and launches
the Universe into the hot Big-Bang era, it should be treated as a part of the complete
inflation theory. However, unlike the inflationary epoch when the physics is almost determined
by the inflationary potential alone, the reheating period is quite complicated
and involves detailed dynamical properties of the model.
While a detailed description of the reheating process is an interesting topic in itself,
our analysis here is motivated by the imprint of the post-inflationary evolution
of the inflaton on primordial fluctuations through the number of $e$-folds, which are sensitive to the mechanism for reheating.
The on-going measurements of CMB observables with increasing precision
are beginning to impose nontrivial constraints on the reheating scenario within a given model of inflation~\cite{DaiReheat}.
One special feature of the GUT models we study is that the inflaton potential changes dramatically
in the post-inflationary era, due to GUT symmetry breaking.
As a result, the original quadratic or exponentially flat inflationary potential
changes into a quartic monomial after the end of inflation, making the Universe effectively
radiation-dominated. The moment of this effective radiation domination can be quite well
determined in our models, and is well before the onset of the reheating process.
In consequence, the number of $e$-folds and the spectral index
in our models are essentially independent of the reheating temperature,
and can be determined relatively precisely.

This paper is organized as follows. In Section\,\ref{sec:2},
we construct models of Higgs inflation in no-scale SUSY GUTs
with the flipped SU(5) and Pati-Salam gauge groups, respectively.
In Section\,\ref{sec:3}, we first analyze the predictions from our no-scale GUT
Higgs inflation models for the scalar tilt $n_s^{}$ and the tensor-to-scalar ratio $\,r\,$.\,
We then study the reheating process after Higgs inflation in these models,
and compute their prediction for the numbers of $e$-folds during inflation.
We further discuss the issue of gravitino production after inflation in these models.
Finally, we draw our conclusions in Section\,\ref{sec:4}.

\vspace*{3mm}
\section{Higgs Inflation in No-Scale Supersymmetric GUTs}
\label{sec:2}
\vspace*{1mm}

 In this Section, generalizing our previous no-scale SU(5) GUT model construction\,\cite{EHX2014},
we construct new models of Higgs inflation in no-scale GUTs with other GUT groups.
In particular, we choose as two concrete realizations the flipped SU(5) GUT~\cite{Antoniadis1987231,Ellis19881}
and the Pati-Salam SU(4)$\,\otimes\,$SU(2)$_L^{}\,\otimes\,$SU(2)$_R^{}$ GUT~\cite{PatiSalam}.
These GUT groups are both major alternatives to the minimal SU(5) GUT,
and can readily be embedded into the SO(10) GUT or accommodated in string theory.

\vspace*{1mm}
\subsection{Higgs Inflation in the Flipped SU(5) GUT}
\label{sec:2.1}
%\vspace*{1.5mm}

The flipped SU(5) GUT has the gauge group SU(5)$\,\otimes\,$U(1) at the GUT scale\,\cite{Antoniadis1987231,Ellis19881},
and thus is only a partial unification. However, this framework has a number of attractive features.
Firstly, it can naturally split the masses of super-heavy colored Higgs bosons
and TeV-scale electroweak Higgs bosons. Secondly, the problematic dimension-5 operator
that causes rapid proton decay is absent in this theory.
Thirdly, the construction of this model does not require Higgs fields in an adjoint representation,
and so can be embedded  easily into perturbative string theory.
In addition, it is easy to incorporate a singlet modulus field $\,T\,$ that may arise from string compactification.

To realize Higgs inflation in the flipped SU(5) GUT,
we need only the minimal field content, namely
a pair of Higgs fields $(G,\,\ob{G})$ in the
$(\mathbf{10},1)$ and $(\mathbf{\overline{10}},-1)$\, representations of SU(5)$\,\otimes\,$U(1)
that are responsible for the GUT symmetry breaking,
and a pair of Higgs fields $(H,\,\ob{H})$ in the
$(\mathbf{5},-2)$ and $(\mathbf{\bar 5},2)$ representations that are responsible for SM symmetry breaking,\, respectively.
We express the components of the GUT Higgs multiplets as follows,
\beqa
\label{FSU5Higgs}
  \begin{aligned}
 &G=\bgp
    0 & d_{G3}^c & -d_{G2}^c & d_{G1} & u_{G1} \\[1mm]
      & 0        & d_{G1}^c  & d_{G2} & u_{G2} \\[1mm]
      &          & 0         & d_{G3} & u_{G3} \\[1mm]
      &          &           & 0 &  \nu_{G}^c  \\[1mm]
      &          &           &        & 0
    \edp \!,~~~
 &&H=\bgp H_{c} \\[1.5mm] H_u
     \edp\!,~~~
 &&\ob{H}=\bgp \ob{H}_{c} \\[1.5mm] \wt{H}_d
     \edp\!,~~~~~
  \end{aligned}
\eeqa
and similarly for $\ob{G}$.

As in \cite{EHX2014}, we introduce the following deformed no-scale K\"ahler potential:
\beqa
\label{KahlerPotential}
\kp\,=
-3\log\bigg[T+T^*-\FR{1}{3}|\Phi_j^{}|^2+\FR{\zeta}{3}\big(H\ob{H}+\text{h.c.}\big)\bigg],
\eeqa
where for convenience we have used units in which the reduced Planck mass
$\,\MP \,(\simeq 2.4\!\times\! 10^{18}$\,GeV) is unity, i.e., $\,\MP =1$.\,
We use $\,\Phi_j^{}=(G,\ob{G}, H,\ob{H},\cdots)$\,
to denote the chiral fields,
where the dots represent fermions that are irrelevant to inflation model building.
Also, we adopt the shorthand notations,
$\,|G|^2\equiv \fr{1}{2}\Tr\!(G^\dag G)$\, and
$\,|H|^2\equiv H^\dag H$,\, etc.
The $\zeta$-term in (\ref{KahlerPotential}) is a natural and slight deformation of
the standard no-scale supergravity K\"ahler potential, and in the following we will allow $\,\zeta\,$
to vary between \,0\, and \,1\,.\,
When $\,\zeta=0$\,,\, the simple no-scale K\"ahler potential is recovered,
and the resultant inflation model has the same predictions as the original Higgs inflation.
On the other hand, when $\,\zeta=1$\,,\, the K\"ahler potential has a shift symmetry
$(H,\,\ob{H})\to (H\!+\!a,\,\ob{H}\!+\!a)$\, with constant $a$,\,
and the resultant inflationary model has a quadratic potential.
It is notable that the shift symmetry is absent for most of our parameter choices,
which is quite different from many models that incorporate supergravity.

In order to discuss the superpotential $\,\sp\,$ of the model,
we first write down the following most general terms up to dimension 4:
\bge
\label{SuperP_Flipped}
  \sp \,
 = -M G\ob{G}-m H\ob{H}+\lam GGH+\bar\lam \ob{G}\ob{G}\ob{H}
   +\al (G\ob{G})^2\!+\be (H\ob{H})^2\!+\ga (G\ob{G})(H\ob{H})\,,~~~~~
\ede
where $\,G\ob{G}\equiv \fr{1}{2}G_{ij}^{}\ob{G}^{ij}$,\,
$GGH\equiv \fr{1}{4}\ep^{ijk\ell m}G_{ij}G_{k\ell}H_m$,\,
and similarly for $\,\ob{G}\ob{G}\ob{H}$.\,
Each of the terms in (\ref{SuperP_Flipped}) is important for our model:
(i)~the $M$ and $\al$ terms collaborate
to break SU(5); (ii).~the $m$ and $\ga$ terms enable the electroweak Higgs doublets
$H_u^{}$ and $H_d^{}$ to be light;
(iii)~the $\lam$ and $\bar\lam$ terms make the colored Higgs fields heavy; and
(iv)~the $\be$ term is important for obtaining a flat inflaton potential during inflation.

To understand these points more clearly, we consider the $F$-term scalar potential:
\beqa
\label{ScalarPotential}
  V(\Phi)\,=\,
  e^{\mathcal{G}}\bigg(\kp^{-1}_{IJ}\FR{\pd\mathcal{G}}{\pd\Phi_I^{}}
  \FR{\pd\mathcal{G}}{\pd\Phi_J^*}-3\bigg)  ,~~~~~
\eeqa
where we use $\,\Phi_I^{}=(T,\,\Phi_i^{})$\,
to denote the modulus $\,T\,$ and other multiplets
$\,\Phi_i^{}=(G,\ob{G},H,\ob{H},\cdots)$\, collectively. In the above,
$\,\mathcal{G} \equiv \kp+\log |\sp|^2$\, and $\,\kp_{IJ}^{-1}$\, is the inverse of the
K\"ahler metric, $\,\kp^{IJ}=\pd^2\kp/\pd\Phi_I^{}\pd\Phi_J^*$\,.\,
As done previously, we assume the modulus $\,T\,$ is stabilized at
$\,\la T\ra=\la T^*\ra=\frac{1}{2}$\, by some high scale physics (see
\cite{Ellis2013_Avatar,HeterModStab2013} for discussions of this point).
Then, at low energies, the $F$-term scalar potential involving $(G,\ob{G})$ becomes
\beqa
  V(G) \,=\, 2G\ob{G}(M-2\al G\ob{G})^2 \, .~~~~~~
\eeqa
Hence, we have a GUT symmetry-breaking vacuum with
$\,\la G\ob{G}\ra=M/(2\al)$\,.\,
Using SU(5) symmetry, we can rotate $\,G\,$ and $\,\ob{G}\,$ such that,
$\,\la \nu_G^c\ra=\la \bar\nu_G^c\ra\equiv v_G=\sqrt{M/2\al}$\,,\,
and all the other components vanish.
At the same time, in order to make the electroweak Higgs doublets $(H_u^{},\,H_d^{})$ light,
we impose the condition $\,m=\ga v_G^2$\,.\,
Then, the $\,\lam\,$ and $\,\bar\lam\,$ terms generate masses for the colored Higgs fields
$\,(H_c^{},\,\ob{H}_c^{})$\,
via the following terms in the scalar potential:
\bge
  V\supset 4\lam^2v_G^2|H_c|^2+4\bar\lam^2v_G^2|\ob{H}_c|^2 \, .~~~~~~
\ede
Supersymmetric GUT unification of the gauge couplings implies that
$\,v_G^{}\simeq 2\times 10^{16}$~GeV,\,
so the colored Higgs fields can be heavy, with masses $\,M_{H_c}^{}=2\lam v_G^{}\,$.
This is a significant advantage over the minimal no-scale SU(5) GUT inflationary model
studied in~\cite{EHX2014}.

 Next, we exploit the no-scale structure of the K\"ahler potential to analyse the inflation potential.
 We identify the $D$-flat component $\,\hat h=|H_u^0|+|H_d^0|$\, as the inflaton,
 which has a value around the Planck scale $\MP$ during inflation.
 This provides a large positive contribution to the effective mass of $G$,\,
 leading to minimization of the potential at $\,G=0$\,.\,
 Inspecting the superpotential, we find that
 this happens when $\,\la H\ob{H}\ra>M/\ga\,$,\,
 which is always satisfied during inflation.
 Hence, we can just set $\,G=0\,$ for the inflation analysis.
 In addition, we can set all the other components of $\,(H,\,\ob{H})\,$
  to zero except for the inflaton field $\,\hat h\,$,\,
  provided that the potential is indeed minimized
  when these components vanish during inflation.
  Checking this stability of inflation trajectory is important, and was done in \cite{EHX2014}.
  We can therefore write the scalar potential
  as a function of the inflaton $\,\hat h\,$ alone,
  and it takes the following simple form:
\beqa
\label{InfPotential}
V(h) \,=\,
\FR{~\big(1-\fr{\be}{2m}\hat h^2\big)^2m^2\hat h^2~}
   {2\big(1-\fr{1-\zeta}{6}\hat h^2\big)^2}\,.
\eeqa
As in \cite{EHX2014}, we impose the condition
$\,\be = \frac{1}{3}(1-\zeta)m\,$
to remove the singularity of $\,V(\hat h)\,$ at
$\,\hat h^2=6/(1-\zeta)$,\,
which would otherwise lead to an exponentially steep potential.
Under this assumption, we derive the following Lagrangian for the inflaton
$\,\hat h\,$, with a non-minimal kinetic term and a quadratic potential:
\beqa
  \ld[\hat h] \,=\,
  \FR{1-\fr{\zeta(1-\zeta)}{6}\hat h^2}{~2\big(1-\fr{1-\zeta}{6}\hat h^2\big)^2~}
  (\pd_\mu\hat h)^2-\FR{1}{2}m^2\hat h^2 \, .
\eeqa
In order to apply the standard slow-roll formalism,
in which the first two slow-roll parameters $\,\ep\,$ and $\,\eta\,$ are obtained
from the inflation potential $\,V(h)\,$ via
\beqa
\ep \,=\, \frac{\,M_{\text{P}}^2\,}{2}\!\(\!\frac{\,V_h^{\prime}\,}{V}\!\)^{\!\!2},
~~~~~~
\eta \,=\, M_{\text{P}}^2\frac{\,V_h''\,}{V} \,,
\eeqa
%
%with the reduced Planck mass $\,\MP =1\,$ in the current convention,
where $\,V_h'=\mathrm{d}V/\mathrm{d}h\,$ and
$\,V_h''=\mathrm{d}^2V/\mathrm{d}h^2\,$,\,
with $\,h\,$ the canonically-normalized inflaton field.
We find that the field $\,h\,$ is connected to $\,\hat h\,$ via
\beqa
  h \,=\,
  \sqrt{6}\,\text{arctanh}\FR{(1\!-\!\zeta)\hat h}
  {~\sqrt{6\!\(\!1-\fr{1}{6}\zeta(1\!-\!\zeta)\hat h^2\)~}~}
  - \sqrt{\FR{6\,\zeta}{\,1\!-\!\zeta\,}\,}\arcsin\!
  \(\!\!\sqrt{\FR{\zeta(1\!-\!\zeta)}{6}}\hat h\!\) \!.~~~~~~~
\eeqa
There are then two interesting limits that can be studied analytically.
One limit is $\,\zeta=0\,$,\,
which gives an exponentially-flat potential in terms of $\,h\,$,\,
and hence the same predictions for the scalar tilt
$\,n_s^{}=1-6\hspace*{0.3mm}\eta+2\hspace*{0.3mm}\ep\,$
and the tensor-to-scalar ratio $\,r=16\hspace*{0.3mm}\ep\,$
as the original models of Higgs inflation\,\cite{HiggsInf} and Starobinsky inflation.
The other limit $\,\zeta=1\,$ yields the quadratic chaotic inflation\,\cite{All}.

\begin{figure}[t]
\centering
\includegraphics[width=0.49\textwidth]{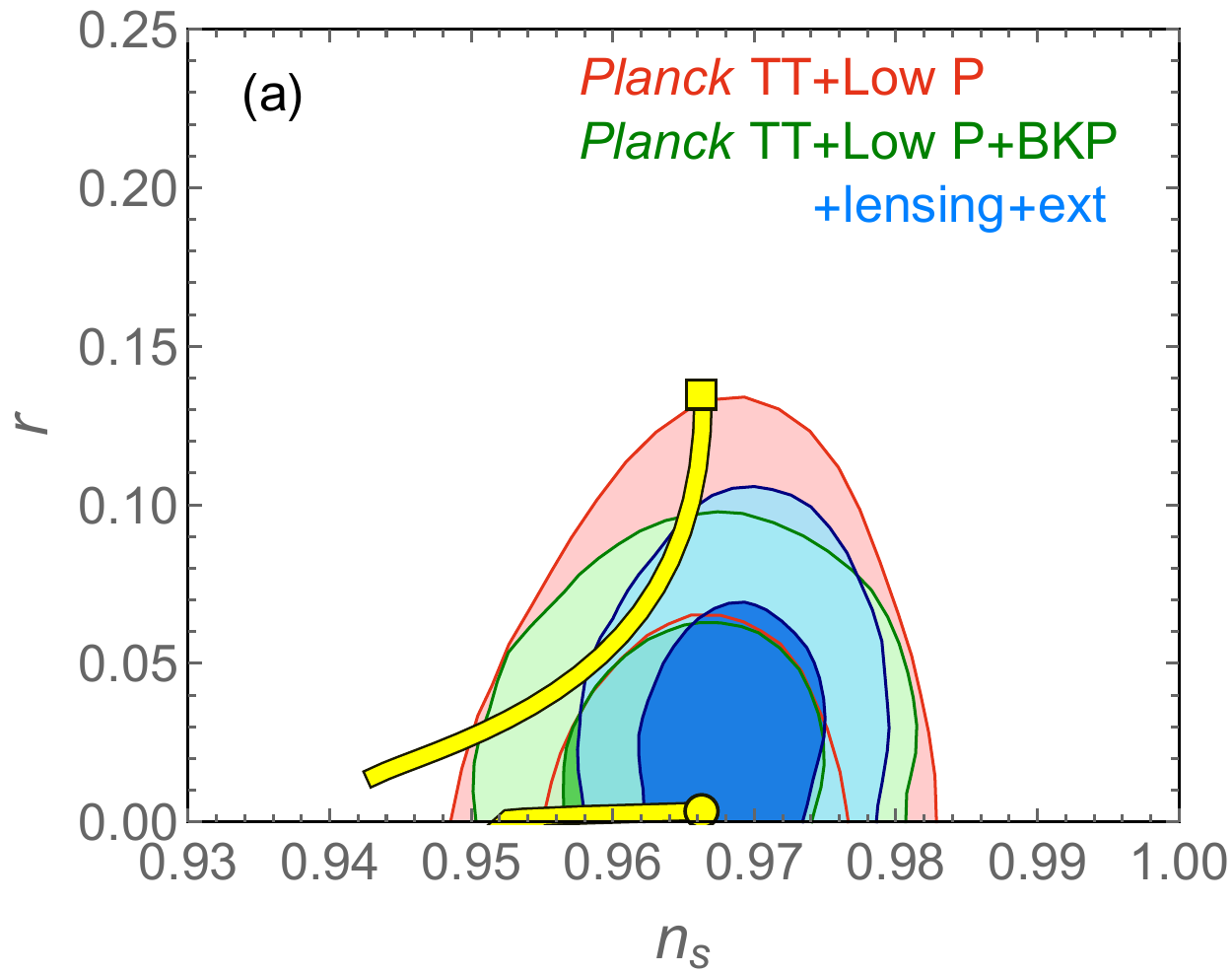}
\includegraphics[width=0.49\textwidth]{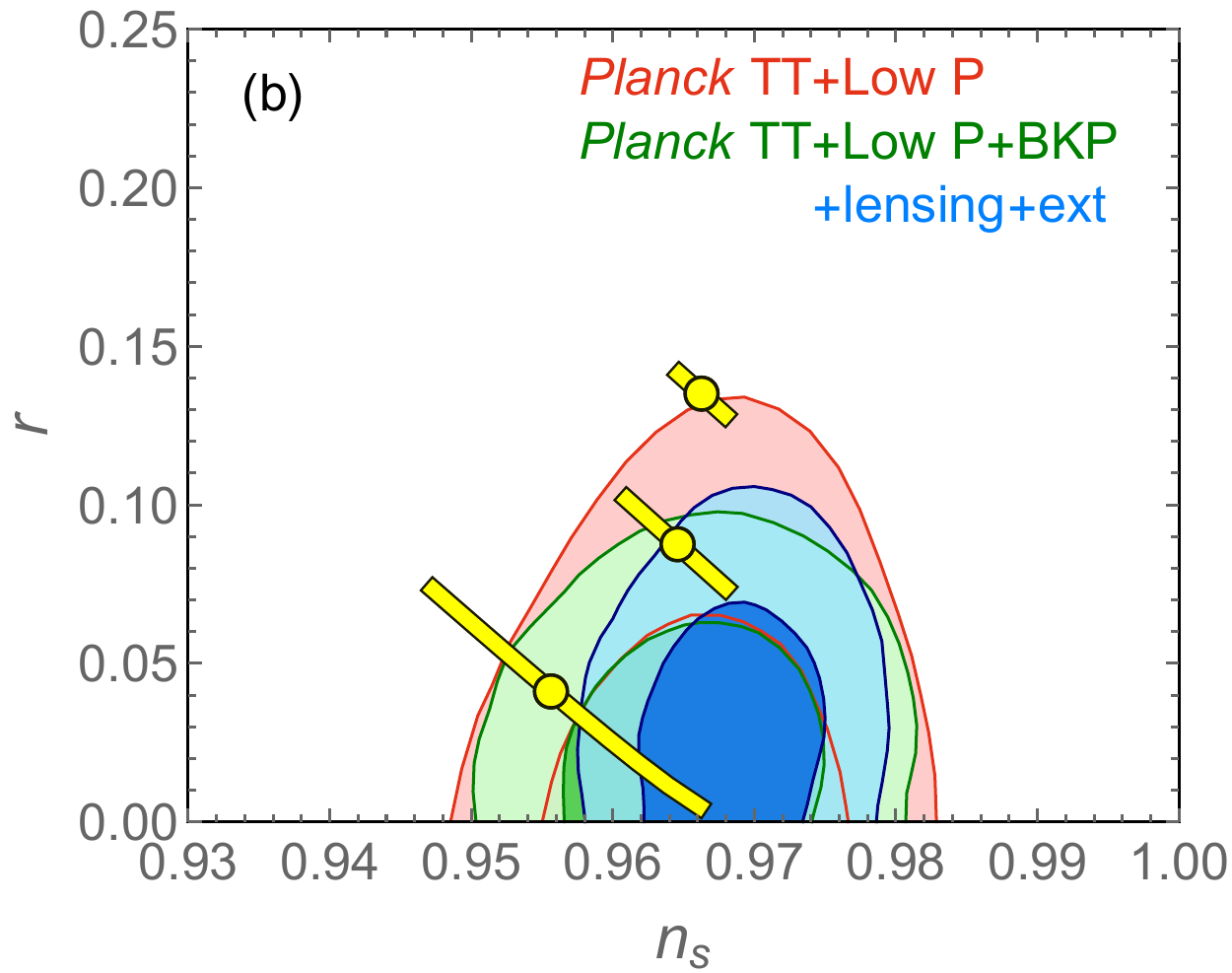}
\vspace*{-2mm}
\caption{\it Predictions from our no-scale GUT models of Higgs inflation
for the scalar tilt $\,n_s^{}$ and tensor-to-scalar ratio $\,r\,$,\,
given the number of $e$-folds $\,N_e=59\,$ from Eq.\eqref{eq:Ne}, and compared
with the 68\% and 95\%\,C.L.\ contours from cosmological observations\,\cite{BKP}.
In plot (a), the condition $\,\be=\frac{1}{3}(1-\zeta)m\,$
is imposed and the round (square) dot corresponds to $\,\zeta=0~(\zeta=1)\,$.\,
The horizontal strip attached to the lower round dot describes
to the effect of varying $\,\zeta \in [0,\,0.1]$\, (from right to left),
while the upper strip attached to the square dot depicts the effect of varying
$\,\zeta\in [0.9,\,1]$\, (from left to right).
In plot (b), the three dots from top to bottom correspond to
$\,\zeta=(1,\,0.98,\,0.95)$\, and $\,\de=0\,$,\, where
$\,\be=\frac{1}{3}(1-\zeta+\de)\,m\,$,\,
and the strip attached to each dot presents the effect of varying
$\,\de\,$ over the range of $\,\pm (1.2\!\times\! 10^{-3})$.}
\label{fignsvsr}
\label{fig:1}
\vspace*{4mm}
\end{figure}

In Fig.\,\ref{fig:1}, we compare the predictions of this model with the recent results
of the Planck Collaboration\,\cite{BKP},
selecting the number of $e$-folds $\,N_e=59\,$, the value
computed in Sec.\,\ref{sec:3}.
In plot (a), we impose the condition $\,\be=\frac{1}{3}(1-\zeta)m\,$,\,
and the round (square) dot corresponds to $\,\zeta=0~(\zeta=1)\,$.\,
The horizontal strip attached to the lower round dot describes
the effect of varying $\,\zeta \in [0,\,0.1]$\, (from right to left),
while the upper strip attached to the square dot presents the effect of varying
$\,\zeta\in [0.9,\,1]$\, (from left to right).
In plot (b), we also analyze the predictions of $(n_s^{},\,r)$
by using a modified condition
$\,\be=\frac{1}{3}(1-\zeta+\de)\hspace*{0.3mm}m$\,,\,
where the parameter $\,\de\,$ varies within
the range of $\,\pm (1.2\!\times\! 10^{-3})$.\,

It is instructive to compare the flipped SU(5) model with the minimal SU(5) GUT model
presented in \cite{EHX2014}. Since the inflation potentials in the two models are identical,
the predictions for $\,(n_s^{},\,r)$\, are the same
when inputting the same number of $e$-folds $N_e$.\,
The same also holds for the predictions of Pati-Salam model, as will be discussed
in the following Section\,\ref{sec:2.2}.
However, it is worthwhile to note some new features
in the case of our flipped SU(5) model of Higgs inflation.

Firstly, we recall that the mass parameter $\,m\,$ in the inflationary potential is fixed
by the Planck normalization of the scalar spectrum to be around $10^{13}$\,GeV.
In the SU(5) model\,\cite{EHX2014}, this is related to the colored Higgs mass
$\,M_{H_c}^{}=\fr{5}{9}m$\, at tree-level, so the colored Higgs boson is rather light,
and hence in tension with the non-trivial constraint from proton stability.
However, in the case of flipped SU(5) we have $\,m=\ga v_G^2\,$,\,
and the colored Higgs boson mass is given by $\,M_{H_c}^{}=2\lam v_G^{}$\,.\,
Hence, the colored Higgs boson is naturally heavy with a mass around $10^{16}$\,GeV.

Secondly, in the case of minimal SU(5), we need to impose a discrete $\mathbb{Z}_2^{}$ symmetry
in order to remove any odd power of adjoint GUT Higgs fields in the superpotential,
which is necessary to produce the desired inflation potential and avoid the colored Higgs mass being
too light. However, in the current flipped SU(5) GUT model
odd powers of the GUT Higgs field $G$ are automatically absent due to the charge assignments,
and there is no need to impose any extra discrete symmetry.

\vspace*{2mm}
\subsection{Higgs Inflation in the Pati-Salam GUT}
\label{sec:2.2}
\vspace*{1.5mm}

Like the flipped SU(5) GUT, the Pati-Salam group
SU(4)$\otimes$SU(2)$_L^{}\otimes$SU(2)$_R^{}$ is also a partial unification,
and can also be embedded readily into SO(10).
However, we will not elaborate on this embedding since the breaking of SO(10)
is irrelevant to the realization of Higgs inflation.
For the current study, we first inspect the relevant field content of the Pati-Salam GUT.
It contains a pair of GUT Higgs multiplets $(G,\,\ob{G})$
in $(\mathbf{4},\mathbf{1},\mathbf{2})$ and
$(\mathbf{\bar 4},\mathbf{1},\mathbf{\bar 2})$ representations, respectively,
together with a $(\mathbf{6},\mathbf{1},\mathbf{1})$ multiplet $D$ and a $(\mathbf{1},\mathbf{2},\mathbf{\bar 2})$ multiplet $H$.
The $(G,\,\ob{G})$ multiplets
can be parameterized as follows:
\beqa
%\begin{aligned}
G=\bgp \bar u_{G1}^c & \bar u_{G2}^c & \bar u_{G3}^c & \bar\nu_G^c \\[1.5mm]
            \bar d_{G1}^c & \bar d_{G2}^c & \bar d_{G3}^c & \bar e_G^c \edp \!,
    ~~~~~~~~
    \ob{G}=\bgp u_{G1}^c & u_{G2}^c & u_{G3}^c & \nu_G^c \\[1.5mm]
            d_{G1}^c & d_{G2}^c & d_{G3}^c & e_G^c \edp \! .~~~~
%\end{aligned}
\eeqa
The $D$ and $H$ fields arise naturally from a $\mathbf{10}$ representation of SO(10)
after its breaking to the Pati-Salam group, namely
$\,\mathbf{10}\to(\mathbf{6},\mathbf{1},\mathbf{1})+(\mathbf{1},\mathbf{2},\mathbf{2})$\,.
The $H$ fields can be parameterized as
\beqa
  H=\bgp H^0_1~ & H_2^+ \\[1.5mm] H_1^-~ & H_2^0 \edp,~~~~~
\eeqa
which splits into the two SU(2)$^{}_L$  Higgs doublets $(H_1^{},\,H_2^{})$ of
the MSSM after the breaking of SU(2)$^{}_R$\,.\,
Finally, the $D$ field can be represented by an antisymmetric tensor of SU(4).

For the superpotential, we choose the following:
\beqa
\sp \,=
-M G\ob{G}-mH^2+\lam DGG+\bar\lam \tilde D\ob{G}\ob{G}+\al(G\ob{G})^2
+\be(H^2)^2+\ga (G\ob{G})H^2,~~~~~
~~~~
\label{SuperP-PT}
\eeqa
where $H^2\equiv\fr{1}{2}\ep^{ij}\ep^{k\ell}H_{ik}^{}H_{j\ell}^{}$,\,
and $\tilde D^{ij}\equiv\ep^{ijk\ell}D_{k\ell}^{}$.\,
The form of the superpotential is similar to that in the previous flipped SU(5) model
(\ref{SuperP_Flipped}). As before, the $M$ and $\al$ terms are responsible
for the GUT breaking. They ensure that $(G,\,\ob{G})$ acquire large expectation values
around the GUT scale, which can be chosen to lie in the $(\bar\nu_G^c,\,\nu_G^c)$ direction,
namely $\la\bar\nu_G^c\ra=\la\nu_G^c\ra=v_G\simeq 2\times\!10^{16}$~GeV.
This breaks SU(4)$\otimes$SU(2)$_R^{}\!\to$\,SU(3)$_C^{}\otimes$U(1)$_{B-L}^{}$.\,
As a result, eight real components in $(\bar u_{Gi}^c,u_{Gi}^c)$ and $(\bar e_G^c,e_G^c)$
are eaten to give masses to gauge bosons corresponding to broken symmetries,
and the other eight components, together with $(\bar\nu_G^c,\nu_G^c)$,
acquire heavy masses $\sim M$.\, At the same time, the
$\,\lam\,$ and $\,\bar\lam\,$ terms ensure that the $(\bar d_G^c,\,d_G^c)$
components and all components of $\,D\,$ also receive masses $\sim M$.

As before, we choose the $D$-flat direction
$\,\hat h= |H_1^0|+|H_2^0|\,$ as the inflaton.
During inflation, only the inflaton $\,\hat h\,$ acquires a large background value around
Planck scale, while all other fields remain at zero.\,
Thus, the scalar potential derived from the above K\"ahler potential and superpotential using
(\ref{ScalarPotential}) is again given by (\ref{InfPotential}), and
the rest of the analysis is the same as for the flipped SU(5) model.

\vspace*{2mm}
\section{Reheating after Higgs Inflation in No-Scale GUTs}
\label{sec:3}

As we have seen, the inflaton in all our GUT models is a linear combination of
the two neutral components in the Higgs doublets of the MSSM.
As such, it couples to various types of matter fields rather strongly,
compared with the gravitational couplings appearing in many typical inflation models.
Consequently, the reheating process can be rather efficient,
and  transfer quickly the energy originally stored in the inflaton potential
to other particles during the classical oscillation of the inflaton field after the inflation.
In particular, when bosons (such as gauge bosons and sfermions) are produced
in this process with a significant accumulation of their number densities,
there could be a period of exponentially fast production of these particles
due to Bose-Einstein statistics, the phenomenon known as stochastic resonance.
The produced particles in this period can be either relativistic or non-relativistic,
depending on their effective masses, which depend on the inflaton background.
The Universe would enter the radiation-dominated era
once most of the inflaton potential energy was released into relativistic particles, and
the collisions of these relativistic particles could then build up
a quasi-thermal equilibrium with reheating temperature $\,\TR\,$.\,
Our next step is to estimate $\,\TR\,$.

The analyses of the reheating process are rather similar for the minimal SU(5) model \cite{EHX2014}
and the flipped SU(5) and Pati-Salam GUT models introduced in Section\,\ref{sec:2}.
We will consider the minimal SU(5) case in the following as an explicit example,
and comment on the differences from the other two models whenever needed.

\vspace*{2mm}
\subsection{The Motion of the Inflaton after Inflation }
\label{sec:3.1}

\begin{figure}
\centering
\includegraphics[width=10cm,height=7.5cm]{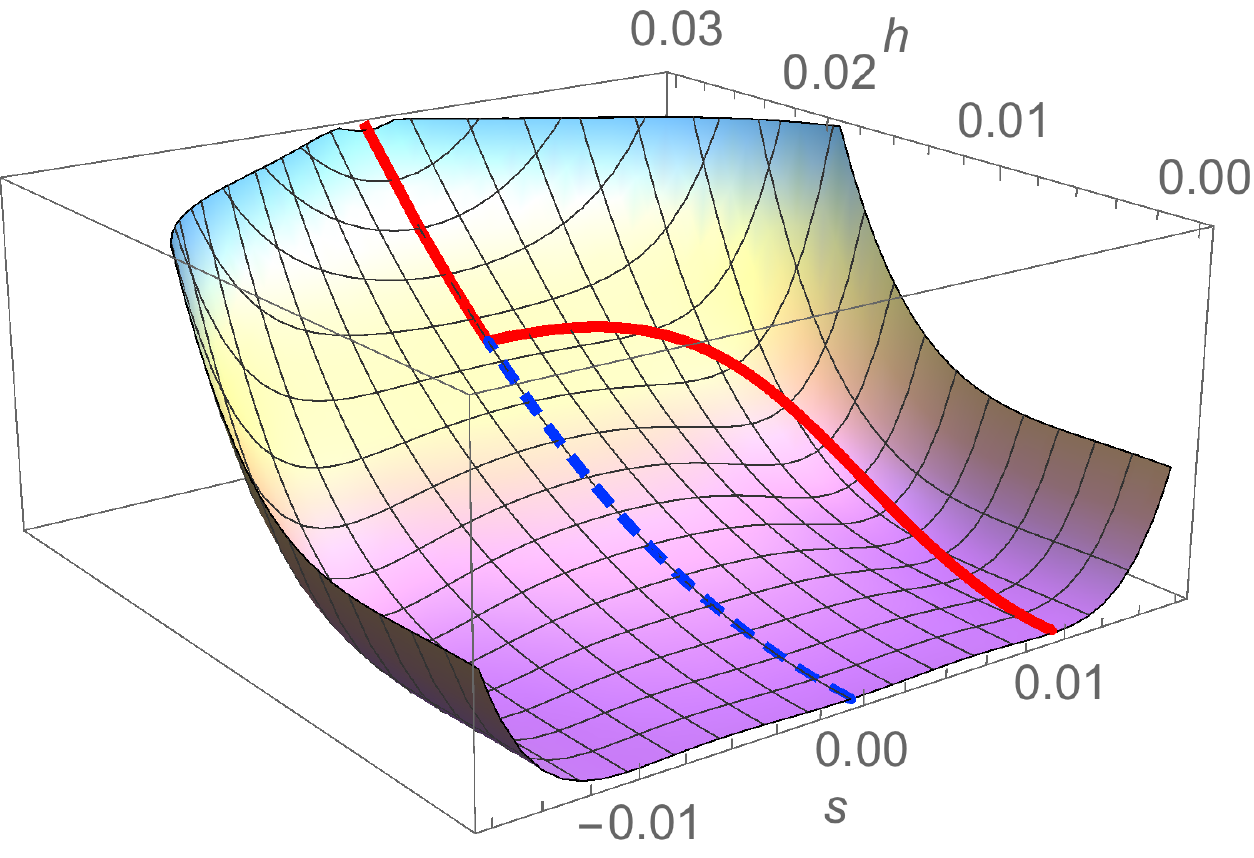}
\caption{{\it Post inflationary trajectory of no-scale Higgs inflation.
This 3-dimensional plot presents the scalar potential $\,V(h,s)\,$ of the flipped SU(5)
model as a function of $\,(h,\,s)\,$ fields.
The red solid curve depicts the trajectory of the inflaton
before and after passing the branch point. The blue dashed curve denotes an (imagined)
continued path under $\,s=0$\,.}}
\label{Contour}
\label{fig:2}
\end{figure}

Inflation ends when the slow-roll parameter $\,\ep\,$ reaches unity,
which happens around $\,h\sim\Mp$.\,
The inflaton then starts to oscillate around the minimum of the potential
at $\,h=0\,$,\, with decreasing amplitude due to the cosmic expansion.
At the first stage of this damped oscillation, the potential is well described
by a quadratic function for all $\,\zeta\in [0,1]\,$:\,
\beqa
\label{QuadInfPot}
  V(h) \; \simeq \; \fr{1}{2}M_h^2 h^2,~~~~~
\eeqa
where the mass parameter $\,M_h^{}\simeq 1.4\!\times\! 10^{13}$\,GeV\,.

We introduce a scalar field $s$ that is the order parameter of GUT symmetry breaking.
In the minimal SU(5) model, $\,s=\text{Re}(\chi)\,$, %is the real component of singlet $\,\chi\,$,\,
where $\,\chi\,$ is a component of the adjoint GUT Higgs $\,\Sigma\,$:\,
$\,\Sigma\supset \sqrt{2/15}\,\text{diag}(1,1,1,-3/2,-3/2)\chi$\,,\,
as we defined in~\cite{EHX2014}.
The interesting feature here is that, in the vicinity of $\,h=0$\,,\,
the path of the inflaton deviates from $\,s=0\,$ due to GUT symmetry breaking.
Quantitatively, the scalar trajectory deviates from $\,s=0\,$
at the branch point $\,h_{\text{br}}^{}\,$,\,
where the trajectory $\,s=0\,$ fails to be the local minimum in the $\,s\,$ direction.
The position of $\,h_{\text{br}}^{}\,$ can be found by solving the condition,
\beqa
\label{eq:BRP-cond}
\left. \frac{\pd^2 V(h,s)}{\pd s^2}\right|_{(h,s)=(h_{\text{br}}^{},0)}^{}
\!=\,0\,.~~~~~
\eeqa
In the minimal SU(5) model of \cite{EHX2014}, the scalar potential $\,V(h,s)\,$
is obtained from the following superpotential:
\beqa
\label{eq:W-mini-SU5}
  W\,=\,
  \al H_u^0(v_G^2-\chi^2)H_d^0+\be(H_u^0H_d^0)^2
  -\frac{1}{2}\lam v_G^2\chi^2+\frac{1}{4}\lam\,\chi^4 \,,~~~~~
\eeqa
and the branch point position derived from the condition (\ref{eq:BRP-cond}) is
\beqa
h_\text{br}^{} =\, v_G^{}\sqrt{2-2\frac{\lam}{\al}+2\sqrt{1\!-2\frac{\lam}{\al}\,}\,}\,,
~~~~~
\label{eq:miniSU5-BP}
\eeqa
where the couplings $\,\alpha,\lambda > 0\,$.\,
The existence of a branch point requires
$\,\frac{\lambda}{\alpha}\leqq 1/2\,$.\,
%$\,\al\geqq 2\lam$\,.\,
In the minimal SU(5) model, $\,\al\simeq 0.06\,$ is fixed
by the amplitude of the curvature perturbation, so we require $\,\lam\leqq 0.03$\,,\,
which is easily satisfied since $\,\lam\,$ is basically a free parameter at this stage.
It is important to note that the parameters above are couplings in the minimal SU(5) model\,\cite{EHX2014},
and differ from the couplings defined for the new models in our previous section.

In the flipped SU(5) model presented in Section \ref{sec:2}
the location of the branch point is
\beqa
h_{\text{br}}^{} =\, 2v_G^{}
\sqrt{1\!-2\frac{\al}{\ga}+\!\sqrt{1\!-4\frac{\al}{\ga}\,}\,}\,,
  ~~~~~
\label{eq:flip-SU5-BP}
\eeqa
where the couplings $\,\alpha,\gamma > 0\,$,\,
and the existence of a branch point requires,
$\,\frac{\alpha}{\gamma}\leqq 1/4\,$ in this case.\,
%$\,\ga\geqq 4\al$\,.\,
Here both $\,\al\,$ and $\,\ga\,$ are free parameters, so they can easily satisfy this condition.

Similarly, we find that the branch point in the Pati-Salam model is given by
\beqa
  h_{\text{br}}^{} =\, 2v_G^{}
  \sqrt{1\!+2\frac{\al}{\ga}+\!\sqrt{1\!+4\frac{\al}{\ga}\,}\,} \,,
  ~~~~~
\label{eq:PT-BP}
\eeqa
which differs from the corresponding expression \eqref{eq:flip-SU5-BP}
in the flipped SU(5) model, due to our choice of couplings in the superpotentials
\eqref{SuperP_Flipped} and \eqref{SuperP-PT}.
Since $\,\alpha,\gamma >0\,$ in this model, the branch point
\eqref{eq:PT-BP} is always present.
From Eqs.\,\eqref{eq:miniSU5-BP}-\eqref{eq:PT-BP},
we further note that the position of the branch point $h_{\text{br}}^{}$ is
around $\,h_{\text{br}}^{}\sim \order{10^{-2}\Mp}$.\,
This is much smaller than the value of the inflaton field at the end of inflation,
as set by the condition $\ep=1$ to be $\,\order{\Mp}$\,.
Hence, the stability of the inflaton trajectory is not affected
by the appearance of the branch point.

In all three models, for $\,h>h_\text{br}^{}$,\,
the inflaton trajectory stays at the local minimum $\,s=0$\,
and the scalar potential in $h$ direction is given by Eq.\,\eqref{QuadInfPot},
whereas for $\,h<h_\text{br}^{}\,$\,
the local minimum of $\,s\,$ shifts away from $\,0\,$, and gradually increases to
$\,s=\vG\simeq 2\!\times\! 10^{16}$\,GeV\, at $\,h=0\,$,
so the corresponding scalar potential deviates from Eq.\,(\ref{QuadInfPot}).
As a result, during the first stage of the post-inflationary oscillation
when the oscillation amplitude of the inflaton is much greater than
$\,h_{\text{br}}^{}$,\, the motion of the inflaton resembles oscillation
in the quadratic potential (\ref{QuadInfPot}).
However, as the amplitude damps, the inflaton trajectory
has $\,s\neq 0$\,, due to the GUT symmetry breaking.
When $\,h\leqq h_\text{br}^{}$,\, the inflaton continues to oscillate
around its local minimum $\,h=0\,$,\, but with $\,s=\vG\,$.\,
The potential $V(h)$ at this stage will be different from (\ref{QuadInfPot}).

The important point here is that the potential becomes very flat when $\,h\,$ is small
and $s\simeq \vG$.
In particular, the Taylor expansion of $\,V(h)\,$ around $\,h=0\,$
does not have a quadratic term $\propto h^2$.
The absence of a mass term can be readily understood.
At the global minimum, the mass of the inflaton (namely the MSSM Higgs boson)
lies at the weak scale, which is much smaller than the scales of inflaton and reheating
under consideration. In general, $\,V(h)\,$ is a complicated function of $\,h\,$.\,
But for our purpose, it is a very good approximation to fit this potential
by a quartic monomial:
\beqa
\label{Vfit}
  V_\text{fit}^{}(h)\,=\,\FR{M_h^2}{\,2h_\text{br}^2\,}h^4 , ~~~~~~(h<h_\text{br}^{}).
\eeqa
In Fig.\,\ref{fig:2}, we present a three-dimensional picture of the scalar potential $\,V(h,s)\,$
in the flipped SU(5) model, where we have the sample inputs
$(\alpha,\,\gamma)=(0.06,\,0.33)$.\,
The red solid curve describes the trajectory of the inflaton
before and after passing the branch point.
As a reference, the blue dashed curve depicts an (imagined)
continued path under $\,s=0$\,.
The potentials in the Pati-Salam model and the minimal SU(5) are quite similar to
that of the flipped SU(5) model, as long as the couplings are chosen such that
the corresponding branch point exists.

Before considering particle production and reheating, we first study the motion of the inflaton
after inflationary epoch, switching off all interactions.
As discussed above, the inflaton $\,h\,$ undergoes a period of oscillations
with decreasing amplitude, initially in a quadratic potential when the amplitude $\,A_h^{}\,$
is larger than the branch point $\,h_{\text{br}}^{}\,$,\, and then via
an effective quartic potential when $\,A_h<h_{\text{br}}^{}\,$.\,
The motion of the inflaton is governed by the following equations:
\beqa
\begin{aligned}
  & 3\Big(\!\FR{\dot a}{a}\!\Big)^2 =\, \FR{1}{2}\dot h^2+V(h)\,,~~~
  \\[1.5mm]
  & \ddot h+3\FR{\dot a}{a}\dot h+V'(h) =\, 0 \,.~~~
\end{aligned}
\eeqa
During the first stage governed by a quadratic potential,
the amplitude of the oscillating solution for $\,h\,$ decreases rapidly
during the first few oscillations.
In fact, the amplitude decreases so fast that after just one oscillation
the amplitude $\,A_h^{}(t)\,$ reduces to $\,0.04\Mp$,\,
already reaching the branch point $\,h_\text{br}^{}\,$ for typical parameter choices.
Hence, soon after inflation the Universe enters the second stage of oscillation
governed by a quartic potential.

During this stage, the oscillation of the inflaton around the local minimum
$\,h=0\,$ and $\,s=\vG\,$\, is no longer harmonic.
For our purpose, it is a good enough approximation to describe
the motion of the inflaton as
\begin{equation}
h(t) \,= A_h^{}(t)\sin\!\left[\int^t\!\!\di t'\omega_h^{}(t')\right]\! ,~~~
\label{omega}
\end{equation}
with amplitude $\,A_h^{}(t)\,$ and frequency $\,\omega_h^{}(t)\,$
that vary slowly with time. Ignoring the cosmological expansion for the moment,
it is easy to find the relation between $\,A_h^{}\,$ and $\,\omega_h^{}\,$
from energy conservation:
\beqa
\label{QuarticFrequency}
\omega_h^{-1} =\,
\FR{1}{\pi}\!\int_{-A_h^{}}^{A_h^{}}\!
\FR{\di h}{\,\sqrt{2[V_\text{fit}^{}(A_h^{})-V_\text{fit}^{}(h)]\,}\,}
\,=\, \FR{2\Gamma(5/4)}{\,\sqrt\pi\Gamma(3/4)\,}
\FR{h_\text{br}^{}}{\,M_h^{} A_h^{}\,} \,.~~~~~
\eeqa
As far as the cosmological expansion is concerned,
we note that oscillations in the quartic potential have an effective equation of state $\,p=\rh/3$\,,\,
which implies $\,a(t)\propto t^{1/2}$,\, $A_h^{}(t)\propto t^{-1/2}$\,
and $\,\omega_h^{}(t)\propto t^{-1/2}$.\,

Hence, the expansion of the Universe during this period is the same as
in a conventional radiation-dominated universe.

\vspace*{2mm}
\subsection{Particle Production}
\label{sec:3.2}

\begin{figure}
 \centering
 \includegraphics[width=10cm,height=8cm]{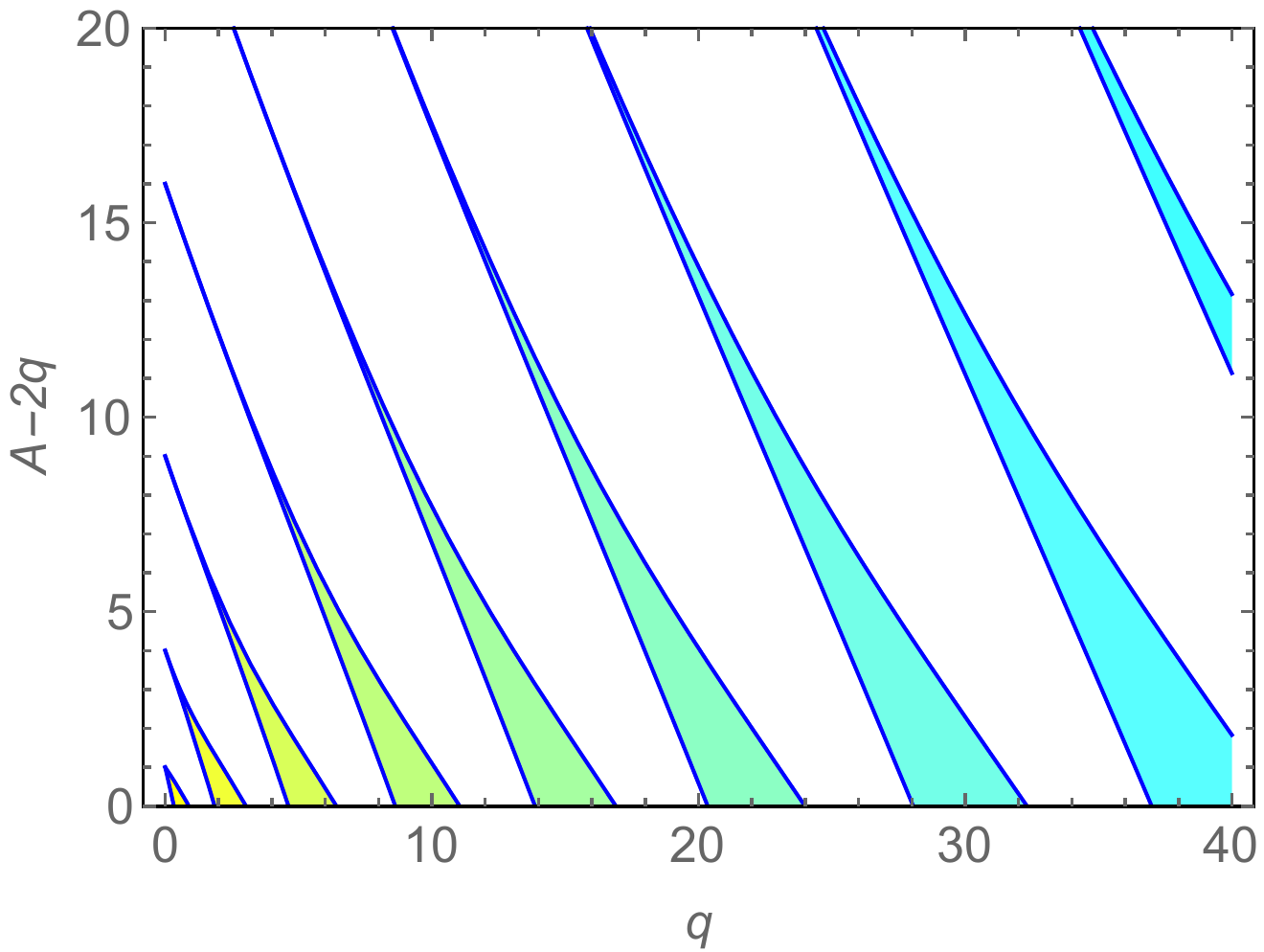}
 \vspace*{-1.5mm}
 \caption{\it Stability-instability chart of the Mathieu equation in the plane of
 $\,q=gA_h^2/(4\omega_h^2)\,$\, versus $\,A-2q=(k^2+M_{\varphi 0}^2)/\omega_h^2\,$.
 The shaded regions represent instability bands, where the exponentially-amplifying solutions are
 located, whereas the conventional perturbative solutions lie in the unshaded regions.}
   \label{fig:3}
   \vspace*{2mm}
 \end{figure}

We now consider interactions and particle production.
The oscillating inflaton field $\,h\,$ may decay perturbatively
to all particles it couples to, so long as this decay is kinematically allowed.
However, it turns out that non-perturbative resonant decays can be more important than perturbative decays
in certain cases \cite{Brandenberger1,Dolgov,Brandenberger2,Linde1994Reheat,Linde1997Reheat}.
To see this point more explicitly, we consider all possible channels for inflaton decays into
bosonic final states, which include gauge bosons, sfermions, $\mathbf{24}$ GUT Higgs bosons $\Sigma$,
and finally the two MSSM Higgs doublets $H_{u,d}^{}$ themselves.\,
We can formulate the production of these particles in the standard way,
treating the inflaton and the Friedman-Robertson-Walker metric as backgrounds, and
studying the evolution of the quantum fluctuations of all decay products. We can write the equation of motion for a given quantum field
$\,\varphi\,$  as follows:
\beqa
\bigg[\FR{\pd^2}{\pd t^2}+3\FR{\dot a}{a}\FR{\pd}{\pd t}-\FR{1}{a^2}\FR{\pd^2}{\pd x^i\pd x^i}+M_\varphi^2(h,a)\bigg]\varphi(t,\mathbf{x}) \,=\, 0 \,,~~~~
\eeqa
where $\,M_\varphi^2(h,a)\,$ is the effective mass of the
quantum field $\,\varphi\,$, which depends on the value of the background inflaton field and the metric.
The contribution from the background metric is negligible in most cases,
so we make a mode decomposition of $\,\varphi(t,\mathbf{x})\,$ as follows:
\beqa
\varphi(t,\mathbf{x})\,=\int\!\!\FR{\di^3k}{(2\pi)^3}
\bigg[a_k^{}\varphi_k^{}(t)e^{+\ii\mathbf{k}\cdot\mathbf{x}}
     +a_k^\dag\varphi_k^*(t)e^{-\ii\mathbf{k}\cdot\mathbf{x}}\bigg],
     ~~~
\eeqa
where the mode $\,\varphi_k^{}(t)\,$ satisfies the following equation:
\beqa
\label{ModeEquation}
\ddot\varphi_k^{}+3\FR{\dot a}{a}\dot\varphi_k^{}
+\bigg[\FR{\mathbf{k}^2}{a^2}+M_\varphi^2(h,a)\bigg]\varphi_k^{} \,=\,0 \,.
~~~~~~
\eeqa
We can infer the number of particles $\,n_k^{}\,$ created by
the mode $\,\varphi_k^{}\,$  by dividing the total energy stored in this mode $\,E_k^{}=\fr{1}{2}(|\dot\varphi_k^{}|^2+\ep_k^2|\varphi_k^{}|^2)$\,
by the energy of each particle,
$\,\ep_k^{}=(\mathbf{k}/a)^2+M_\varphi^2$\,,\,
yielding
\beqa
  n_k^{} \,=\, \FR{1}{\,2\ep_k^{}\,}\!
  \(|\dot\varphi_k^{}|^2+\ep_k^2|\varphi_k^{}|^2\) .~~~~~~
\eeqa
To obtain an intuitive picture how the resonance would happen,
it is instructive to consider an idealised case in which the expansion of the Universe
could be ignored. In this case, we would have $\,a=1\,$,\,
and the background inflaton $\,h\,$ would oscillate simply as
$\,h(t)=A_h^{}\sin \omega_h^{} t$\,.\,
Then, if the quantum field $\,\varphi\,$ couples to the inflaton $\,h\,$
through an interaction term
$\,\fr{1}{2}g^2 h^2\varphi^2$\, with coupling constant $\,g\,$,\,
the effective mass of $\,\varphi\,$ can be written as
$\,M_\varphi^2 = M_{\varphi0}^2+g^2A_h^2\sin^2\omega_h^{} t$\,,\,
where $\,M_{\varphi 0}^{}\,$ is the mass of $\,\varphi\,$
in the absence of the background inflaton field.
In consequence, the mode equation for $\,\varphi_k^{}\,$
reduces to the well-known Mathieu equation:
\beqa
\label{MathieuEquation}
\FR{\di^2\varphi(\xi)}{\di\xi^2}+(A-2q\cos 2\xi)\varphi(\xi) \,=\,0 \,,~~~~~~
\eeqa
where $\,\xi = \omega_h^{} t$,\,
$A = (\mathbf{k}^2+M_{\varphi 0}^2+\fr{1}{2}g^2A_h^2)/\omega_h^2$\,,\,
and $\,q = g^2A_h^2/(4\omega_h^2)$.\,
The behaviour of a solution to this equation depends on the parameters $A$ and $\,q\,$.\,
As shown in Fig.\,\ref{fig:3}, the plane $(A,\,q)$ can be divided into stable and unstable regions,
where the stable regions (unshaded) correspond to ordinary oscillating solutions,
and the unstable regions (shaded) correspond to exponentially-amplified solutions.
It is the latter that give rise to resonant production of particles.

In a more realistic case, the large amplitude of the first few oscillations
and the cosmic expansion of the Universe make the situation more complicated.
It is possible, in general, that the parameters $(A,\,q)$ scan a large number of
instability bands within a few oscillations. As a result, the resonant production
of particles can behave in a stochastic manner.
For this reason, this period of particle production is termed stochastic resonance.

With this in mind, we now consider the couplings
between the inflaton field $\,h\,$ and various bosons
into which it may decay. We consider first the SU(5) gauge bosons,
which can be parameterized via the following $5\!\times\!5$ matrix:
\beqa
  \mathbf{A}_{\mu}^{} =
  \bgp
  \fr{1}{\sqrt{2}\,}\lam_{\text{SU(3)}}^a G^a_\mu+\!\sqrt{\!\fr{2}{15}}V_{24\mu}^{}
  & X_\mu^{-4/3} & Y_\mu^{-1/3}
  \\[1.5mm]
  X_\mu^{+4/3}
  & \fr{1}{\sqrt{2}\,}W_{3\mu}^{}\!-\!\sqrt{\!\fr{3}{10}}V_{24\mu}
  & \fr{1}{\sqrt{2}\,}(W_{1\mu}^{}\!-\ii W_{2\mu}^{})
  \\[1.5mm]
  Y_\mu^{+1/3} & \fr{1}{\sqrt{2}\,}(W_{1\mu}^{}\!+\ii W_{2\mu}^{}) & -\fr{1}{\sqrt{2}\,}W_{3\mu}^{}\!-\!\sqrt{\!\fr{3}{10}}V_{24\mu}^{}\edp \!.~~~~~~
\eeqa
We can deduce the effective masses for various gauge bosons
in the inflaton background $\,h\,$ directly from the kinetic terms $\,|\D_\mu^{} H_{1,2}^{}|^2$:\,
\beqa
%\begin{aligned}
  M_X^2=M_Y^2=\FR{5}{3}g^2v_G^2+\FR{1}{4}g^2h^2,
  ~~~~~
  M_W^2=\FR{5}{3}M_{V_{24}}^2=\FR{1}{4}g^2h^2,~~~~~~
%\end{aligned}
\eeqa
where $\,g\,$ is the SU(5) gauge coupling at the GUT scale.

We consider next the sfermions in the supersymmetric SU(5) model,
which couple to the background inflaton via Yukawa terms in the superpotential:
\beqa
  W_{\text{Yukawa}}^{}=
  \FR{1}{8}y_u^{}\ep^{ijk\ell m}T_{ij}^{}T_{k\ell}^{}H_{m}^{}
  +y_d^{} T_{ij}^{}\ob\psi_i^{} \ob H_{j}^{} \,.~~~~~~
\eeqa
Here $T_{ij}^{}\,$ is a left-handed multiplet in a $\mathbf{10}$ representation,
and $\,\ob\psi_i^{}\,$ is a left-handed multiplet in a $\mathbf{\ob{5}}$ representation,
which can be parameterized as follows:
\beqa
%\begin{aligned}
T = \bgp 0 & (u^c_3)_L^{} & -(u_2^c)_L^{} & u_L^1 & d_L^1 \\[1.2mm]
     &  0 & (u_1^c)_L^{} & u_L^2 & d_L^2 \\[1.2mm]
     &  & 0 & u_L^3 & d_L^3 \\[1.2mm]
     &  & & 0 & e^+_L \\[1.2mm]
     & & & & 0 \edp \!,
~~~~~
\ob\psi=\bgp d_L^{1c} \\[1.2mm]
d_L^{2c} \\[1.2mm]
d_L^{3c} \\[1.2mm]
e_L^- \\[1.2mm]
\nu_L^{} \edp \!,
%\end{aligned}
\eeqa
where $T_{ij}^{}$ is antisymmetric, and for simplicity we suppress all the flavor indices.
The effective masses extracted from the scalar potential are
\beqa
%\begin{aligned}
  M_{uL}^2=\FR{1}{4}(y_u^2+y_d^2)h^2 ,~~~
  M_{uR}^2=\FR{1}{4}y_u^2 h^2 ,~~~~
  M_{dR}^2=M_{\ell L}^2=M_{\nu R}^2=\FR{1}{4}y_d^2 h^2 .~~~~~~
%\end{aligned}
\eeqa
In the realistic case with three generations of fermions,
the sfermion mass spectrum would be obtained by diagonalization in flavor space
and will be more complicated in general, and depend on model assumptions.
However, for the sake of illustration we do not elaborate on details of the flavor structure.

The inflaton $\,h\,$ also gives effective masses to the $\,\Sigma_{ij}^{}$ $(i,j=4,5)$\,
 components of the $\mathbf{24}$ GUT Higgs multiplet $\Sigma$\,, as follows:\,
\beqa
  M_\Sigma^2 \,=\, \lam^2v_G^4+\FR{1}{4}\al^2h^4 \, ,
\eeqa
where both $\lam$ and $\al$ have mass dimension $-1$,\,
as is clear from Eq.\,\eqref{eq:W-mini-SU5}.
Finally, the coupling of the inflaton to the other components of the MSSM Higgs doublets
$H_{u}^{}$ and $H_{d}^{}$ can safely be neglected, since the inflaton is moving
in the $D$-flat direction, so its $F$-term couplings to other components in
$H_{u,d}^{}$ are all of higher order and thus are highly suppressed.

The above analysis shows that the effective masses of various bosons
are generally of order $h$ during the era of inflaton oscillation, and
hence are much larger than inflaton mass $\,M_h\sim 10^{13}\,$GeV.
Hence, at this stage the perturbative decay of inflaton is kinematically forbidden,
and the leading channel for energy transfer is through non-perturbative resonances.
Fermion production through perturbative decay may also present,
but is generally subdominant.

An interesting feature of resonant production here is that the oscillation frequency
$\,\omega_h^{}\,$ of the background inflaton $\,h\,$ field is proportional to the amplitude
$\,A_h^{}\,$.\,  In consequence, the $\,q\,$ parameter in the Mathieu equation remains
constant with $\,q=g^2A_h^2/(4\omega_h^2)=\order{1}\!\times\! (gh_\text{br}^{}/M_h)^2$.\,
Recalling $\,h_\text{br}^{}\sim \order{10^{-2}\Mp}$\,
and $\,M_h^{}\sim 10^{13}$\,GeV,
we see that $\,q\,$ is a large number, of $\,\order{10^4}\,$ to $\,\order{10^5}$.\,
Hence a broad resonance can readily appear when $\,A-2q\,$ is small.
Also, we note that $\,A-2q=(\mathbf{k}^2+M_{\varphi 0}^2)/\omega_h^2$\,,\,
so that for a field $\,\varphi\,$ with $\,M_{\varphi 0}^{}=0\,$,\,
small $\,A-2q\,$ is easily achieved for small $\,\mathbf{k}^2$.\,
On the other hand, for heavy particles such as $(X,\,Y)$ gauge bosons
and $\,\Sigma\,$ bosons (including the $s$ field),
the $\,A\!-2q\,$ parameter increases linearly with time,
so the broad resonance is suppressed for these species.
In summary, we see that particle production via broad resonance could be efficient
only for the light gauge bosons $(W,\,V_{24}^{})$ and for sfermions \cite{Lozanov}.

A broad resonance leads to an exponential increase of the number $\,n_k^{}\,$
of particles in a given mode, $\,n_k^{}\propto e^{2\mu_k^{}\omega_h^{}t}$,\,
where $\,\mu_k^{}\,$ is a coefficient of $\,\order{0.1}$\,
that can be determined by solving equation (\ref{ModeEquation})
or the Mathieu equation (\ref{MathieuEquation}) by ignoring the cosmic expansion.
The resultant $\,\mu_k^{}\,$ for $\,A-2q=0\,$ is a rapidly varying function of $\,q\,$,\,
taking values between 0.15 and 0.35.\, The total number of produced particles
can then be inferred by integrating over the particle numbers of all modes.
Since the zero-mode particles have the largest coefficient $\,\mu_{k=0}^{}$\,,\,
it is reasonable to estimate the number of produced particles by considering zero modes only,
in which case we have $\,n\sim\exp(2\mu_0^{}\omega_h^{} t)$\,.

The above analysis ignores processes that can decrease the number of produced particles,
including decays and scattering with other particles.
These processes turn out to be very important, and may destroy the resonant production
of gauge bosons and sfermions, as we illustrate using the $W$ boson as an example.
The decay rate of $\,W^\pm\,$ is of order
$\,\Gamma_W^{}\sim g^2\la M_W^{}\ra$,\, where $\,\la M_W^{}\ra\,$
is the averaged $W$ mass during an oscillation of the inflaton $\,h\,$,\,
which is given by $\,\la M_W\ra=\fr{1}{2}g\la |h|\ra$.\,
At the same time, $W^\pm$ pairs can annihilate through scattering,
with a cross section $\,\si_W^{}\sim\la M_W^{}\ra^{-2}$.\,
As a result, the number density $\,n_W^{}\,$ of $W$ bosons is given by
\bge
\label{nWrate}
  \FR{\di}{\di t}\(a^3n_W^{}\) \,=\,
  a^3\big(2\mu_0^{}\omega_h^{} n_W^{}-\Gamma_W^{} n_W^{}-\si_W^{} n_W^2\big) .
\ede
In the first stage of reheating during which the number density of $\,n_W^{}\,$
is small, the scattering process is rare and its rate is suppressed by $\,n_W^2$.\,
So, we should compare the production rate
$\,2\mu_0^{}\omega_h^{}\,$  in (\ref{nWrate}) with the decay width $\,\Gamma_W^{}\,$
to determine whether $W$ decay can disrupt resonant production.
From Eq.\,(\ref{QuarticFrequency}) and $\,\Gamma_W^{}\sim \fr{1}{2}g^3\la |h|\ra$,\,
and the fact that $\,\la |h|\ra\propto A_h^{}$,\, we see that $\,\omega_h^{}\,$ and
$\,\Gamma_W^{}\,$ have the same dependence on the oscillation amplitude $\,A_h^{}$.\,
Hence, the decay process is a more rapid process than resonant production when
\bge
\label{CoefComp}
  \FR{\,2\mu_0^{}\sqrt{\pi}\,\Gamma(3/4)M_h^{}\,}{2\Gamma(5/4)\,h_{\text{br}}^{}}
  \,\lesssim\, \FR{1}{2}g^3,
\ede
with an $\order{1}$ uncertainty in the coefficient.
Since $\,\mu_0^{} M_h^{}/h_{\text{br}}^{}\sim \order{10^{-4}}\,$
and $\,g\sim\order{0.1}$,\, we see that the decay of the $W$ boson is so quick
that resonant production cannot take place efficiently.
This conclusion is certainly not definitive, since the left-hand-side of (\ref{CoefComp})
is not parametrically small, so some $\order{1}$ uncertainty may alter this picture,
and a period of not-very-efficient resonant production may happen.

Without getting involved in these details, we see that the reheating temperature $\,\TR\,$
of this model may be significantly lower than in conventional SM Higgs inflation,
which is estimated to be around $10^{14}$\,TeV~\cite{Garcia2009_HIreheat,Bezru2009_Reheat},
due to the quartic shape of the effective potential (\ref{Vfit}).
We recall that a low reheating temperature in supersymmetric GUTs
may help to avoid the over-production of gravitinos (as we discuss below)
as well as unwanted topological defects.

Although a more precise estimate of the reheating temperature
depends whether the decays of gauge bosons and sfermions
would disrupt their resonant production, which further depends on numerical details,
it is important and interesting to note that the number of $e$-folds $N_e$
can be determined without ambiguity. This is because $N_e$ in our models does not depend on
the reheating temperature $\TR$,
but is determined by the energy density $\,\rh_{\text{rad}}^{}\,$
when radiation begins to dominate the Universe.
Since the above analysis shows that the scalar potential changes from a quadratic shape
to a quartic one at the branching point $\,h=h_{\text{br}}^{}$,\,
we immediately deduce,
$\,\rh_{\text{rad}}^{} \simeq M_h^2h_{\text{br}}^2$\,,\,
with $\,h_{\text{br}}^{}$\, given by Eqs.\,(\ref{eq:miniSU5-BP}), (\ref{eq:flip-SU5-BP}),
and (\ref{eq:PT-BP}) for the three models, respectively.
For instance, in the minimal SU(5) GUT model, we have
\beqa
\label{rhorad}
\rh_{\text{rad}}^{}\,\simeq\,
M_h^2h_{\text{br}}^2
\,\simeq\, 2\al^2v_G^6\!\(\!1-\frac{\lam}{\al}+\sqrt{1-2\frac{\lam}{\al}\,}\) \!.
\eeqa
Hence, using\,\cite{Planck2013Inf,Planck2015_Inflation},
we compute the number of $e$-folds to be
\beqa
\label{eq:Ne}
  N_e \,\simeq\,
  62+\FR{1}{4}\log\!\bigg(\!\FR{V_{\text{begin}}^{}}{\Mp^4}\!\bigg)
  +\FR{1}{4}\log\!\bigg(\!\FR{V_{\text{begin}}^{}}{\rh_{\text{end}}^{}}\!\bigg)
  +\FR{1}{12}\log\!\bigg(\!\FR{\rh_{\text{rad}}^{}}{\rh_{\text{end}}^{}}\!\bigg)
  -\FR{1}{12}\log g_* \,\simeq\, 59\,,~~~~~~~
\eeqa
where $\,V_{\text{begin}}^{}$ and $\,\rh_{\text{end}}^{}$\,
denote the energy density of the inflaton at the beginning
and the end of observable inflation, respectively, and $\,g^{}_*$\, is the effective number
of degrees before the moment of effective radiation dominance, which is $\,\order{1}\,$
and contributes little to $N_e$\,.\,
Finally, we deduce the number of $e$-folds $\,N_e\simeq 59\,$
by inputting $\,V_{\text{begin}}^{}\simeq \rh_{\text{end}}^{}\simeq
\fr{1}{2}\al^2v_G^4$\, and $\,\rh_{\text{rad}}^{}\simeq 2\al^2v_G^6$\,
with $\,\al\simeq 0.06$\, and $\,\vG\simeq 0.01$\,.
For the flipped SU(5) model and the Pati-Salam model,
the result $\,N_e\simeq 59\,$ also holds well. This is because
$\,\rh_{\text{rad}}^{}\,$ only differs by $\order{1}$ factors among the three models,
and thus has negligible difference when computing $N_e$,\,  due to the
very mild logarithmic dependence $\,\log\rh_{\text{rad}}^{}\,$ suppressed by a
small coefficient $\,\frac{1}{12}\simeq 0.08$\, as in Eq.\,\eqref{eq:Ne}.
In passing, we also note that the estimate of $\,N_e\,$ is not affected
by the accumulated decay products even when they start to dominate the energy density,
because these particles are also light and highly relativistic, and thus the Universe is always
effectively dominated by radiation once the branch point is reached.

In the above, we have mainly presented the explicit analysis for the minimal SU(5) model
\cite{EHX2014}, but it is clear that the two principal conclusions in this section apply also
to the flipped SU(5) and Pati-Salam models given in Section\,\ref{sec:2}.
Namely, (i).~the reheating temperature can be as high as $10^{14}$\,GeV,
but is probably much lower, where the uncertainty is mainly due to
the highly non-perturbative dynamics of the reheating process,
which depends on model details;
(ii).~the number of $e$-folds is determined to be $\,N_e\simeq 59\,$,\,
and is independent of details of the reheating process.
This is because $N_e$ depends only logarithmically on
the scale of effective radiation dominance $\,\rh_{\text{end}}^{}\,$,
and $\,\rh_{\text{end}}^{}\,$ is roughly the GUT symmetry-breaking scale,
which is the same for all three models.

With $N_e$ derived, we can predict the values of scalar tilt $\,n_s^{}\,$
and tensor-to-scalar ratio $\,r\,$ more precisely,
as represented by the yellow strips in Fig.\,\ref{fig:1} (Sec.\,\ref{sec:2.1}), where
the predictions of our models are compared with the latest Planck results in 2015
\cite{BKP,Planck2015_Inflation}.

\vspace*{2mm}
\subsection{Gravitino Production}
\label{sec:3.3}

It is also desirable to study gravitino production in our models, since the ratio
$\,Y_{3/2}^{}\equiv\rh_{3/2}^{}/\rh_r^{}\,$
between the energy densities of the gravitino $\,\rh_{3/2}^{}\,$
and radiation $\,\rh_r^{}\,$ is generally subject to nontrivial constraints
from gravitino production \cite{Khlopov1,BBNgravitino2008}.
The standard calculation of $\,Y_{3/2}^{}\,$, assuming instant decay
of the inflaton and thermalization, yields
\beqa
\label{thermalY32}
Y_{3/2}^{} \,\simeq\,
\FR{\,0.00398\,}{\sqrt{c\,}}\(\!\FR{\Gamma_h^{}}{\Mp^{}}\!\)^{\!\!\frac{1}{2}}
\!\(\!1+0.558\FR{m_{1/2}^2}{\,m_{3/2}^2\,}\!\)e^{-t\,\Gamma_{3/2}} \,,~~~~~~
\eeqa
where the constant $\,c=\order{1}\,$,\,
$\,\Gamma_h^{}\,$ ($\,\Gamma_{3/2}^{}\,$) is the decay rate of  the inflaton (gravitino),  and
$\,(m_{1/2}^{},\,m_{3/2}^{})$\, are the gaugino mass at the GUT scale and the gravitino mass, respectively.
It is also assumed that the effective degrees of freedom during reheating consist of the MSSM fields.
It was shown recently \cite{Gravitino2015} that the above expression is a good estimate
with $\,c=1.2\,$,\, even when we take into account of perturbative gravitino production
prior to thermalization. However, as was shown above,
non-perturbative resonance would probably occur in our models, and gravitino production in the
non-perturbative regime may or may not affect the above standard estimation.

The non-perturbative production of gravitinos can be important in our models.
To estimate the number density $\,n_{3/2}^{}\,$ of gravitinos produced
during a non-perturbative resonance, it is a good approximation to consider the helicity-$\frac{1}{2}$
states only, because they are produced more efficiently than the helicity-$\frac{3}{2}$ states
(which couple to other fields only through gravity).
The resonant production of helicity-$\frac{1}{2}$ states is similar to
that of a spin-$\frac{1}{2}$ fermion, with low-momentum states dominating,
since the instability region is denser at low momentum, as seen in Fig.\,\ref{fig:3}.
In the most efficient scenario, the produced gravitinos occupy all momentum states up to
a ``fermi surface'' $k_f^{}$, while all $\,k>k_f^{}\,$ states are essentially empty.
The physical momentum $\,k_f^{}\,$ cannot exceed the scale of the energy density during reheating,
which is always below $\,\rh_{\text{rad}}^{1/4}\,$ in Eq.\,(\ref{rhorad}).
Therefore, we derive the following upper limit on $\,n_{3/2}^{}\,$ production through non-perturbative
effects:
\beqa
(n_{3/2}^{})_{\text{rad}}^{} \,\lesssim\, \rh_{\text{rad}}^{3/4}\,.\,
\eeqa
The subscript indicates that the number density is evaluated
at the time of effective radiation dominance.

In order to compare this result with the thermal production (\ref{thermalY32}),
we note that the number density scales as $\,a^{-3}\,$ with $\,a\,$ the scale factor,
and the energy density scales as $\,a^{-4}\,$ due to effective radiation dominance.
Hence, at the time of thermalization, namely the time of reheating, the number density
$\,(n_{3/2}^{})_{\text{reh}}^{}\,$ is diluted to
\beqa
n_{3/2}^{}\,\sim\, (\rh_{\text{reh}}^{}/\rh_{\text{rad}}^{})^{3/4}(n_{3/2}^{})_{\text{rad}}^{}
\,\lesssim\, \rh_{\text{reh}^{}}^{3/4} \,\sim\, T_{\text{reh}^{}}^3 \, .~~~~~
\eeqa
Thus, the energy density of gravitinos is
$\,\rho_{3/2}^{}\!\sim\! m_{3/2}^{} T_{\text{reh}}^3$.\,
Since the radiation energy density at the time of reheating is of the order
$\,\rho_{r}^{}\!\sim\! T_{\text{reh}}^{4}$,\, we divide the gravitino energy density
by the radiation energy density, and find that the non-perturbative production
of gravitinos contributes to the ratio $\,Y_{3/2}^{}=\rh_{3/2}^{}/\rh_r^{}\,$
an amount of
$\,{\cal O}(m_{3/2}^{}/T_{\text{reh}}^{})\,$.
This contribution could be made cosmologically acceptable by requiring the gravitino
to be either heavy or very light.
A heavy gravitino with mass
$\,m_{3/2}^{}>\order{10-100}$\,TeV\,
would have decayed into radiation before Big-Bang nucleosynthesis (BBN),
and so would be harmless. In the case of an ultra-light gravitino with mass
$\,m_{\text{3/2}}^{}\ll 1$\,GeV,\,
the bound $\,Y_{3/2}<\order{10^{-14}}$\, can be satisfied
for $\,T_{\text{reh}}^{}\lesssim \order{10^{14}}$\,GeV.

\vspace*{2mm}
\section{Conclusions}
\label{sec:4}
\vspace*{2mm}

Higgs inflation identifies the inflaton field as the observed Higgs boson, and
provides a truly economical approach to realize the cosmic inflation
that could have driven the exponential expansion of the very early Universe
and generated the observed large scale structure.
The desired energy scale of successful inflation lies around
the scale of supersymmetric gauge unification, providing a strong motivatation to embed
Higgs inflation into attractive no-scale supersymmetric GUTs.

In this work, we have extended our previous study on no-scale inflation
in the minimal SU(5)~\cite{EHX2014}
to a class of Higgs inflation models in no-scale supersymmetric GUTs
with different groups, namely the flipped SU(5) and Pati-Salam group,
as presented in Section\,\ref{sec:2}.
The colored Higgs mass $\,M_c\,$ is more flexible in these models, and can be heavier in both
the flipped SU(5) and Pati-Salam models, compared to the minimal SU(5) model.
This helps to remove the tension between GUT models and the proton stability constraint.
Then, in Section\,\ref{sec:3},
we studied systematically the reheating process after Higgs inflation
\`{a} la no-scale GUT. We showed that the number of $e$-folds can be determined without ambiguity,
due to the quartic shape of the scalar potential in the lead-up to the reheating process.
We have derived $\,N_e\simeq 59$,\, which yields predictions for the scalar tilt
and the tensor-to-scalar ratio that are consistent with the current observational limits
as shown in Fig.\,\ref{fig:1},
and will be further tested by more precise measurements of scalar tilt $\,n_s^{}\,$
in the near future. Unlike $\,N_e$,\, we note that the reheating temperature $\,\TR\,$
at which particles were thermalized depends on more details of the reheating process,
in particular the efficiency of resonant production.
Our simple estimate has shown that the reheating temperature $\,\TR\,$ in these models is generally
lower than that in the conventional SM Higgs inflation.
It is desirable to estimate the reheating temperature
more precisely in these models, since $\,\TR\,$ in supergravity inflation models is generally subject
to an important constraint from the gravitino production rate \cite{BBNgravitino2008}.
In order to prevent over-production of gravitinos, a relatively high reheating temperature
would require the gravitino mass to be either very heavy (above $10-100$\,TeV)
or ultra-light (much below $1$\,GeV) \cite{Ferrara2011_SCon},
which would have important implications for supergravity phenomenology.

%\newpage
\vspace*{5mm}
%\addcontentsline{toc}{section}{Acknowledgments\,}
\noindent
{\bf\large Acknowledgements}\\[1.5mm]
 The work of JE was supported in part by the STFC Grant ST/J002798/1,
 and the work of HJH and ZZX was supported in part by the National NSF of China, 
 under grants 11275101 and 11135003. 
 % and by Tsinghua University (under grant 20141081211).

%\paragraph{Acknowledgements.}

%\end{appendix}

%\vspace{10mm}
%\newpage
%

\end{document}